\documentclass[dvipdfm]{elsart}

\usepackage{graphicx, epsfig}
\usepackage{amsmath,amssymb}
\usepackage[latin1]{inputenc}

\newcommand{\bsy}{\boldsymbol} 
\newcommand{\Es}{\mathbb{E}}

\begin{document}

\begin{frontmatter}
\title{Improved Likelihood Inference in Birnbaum--Saunders Regressions}
\author{Artur J.~Lemonte, Silvia L.~P.~Ferrari}
\address{Departamento de Estatística, Universidade de São Paulo,
Rua do Matão, 1010, São Paulo/SP, 05508-090, Brazil}
\author{Francisco Cribari--Neto}
\address{Departamento de Estatística, Universidade Federal de Pernambuco,
Cidade Universitária, Recife/PE, 50740-540, Brazil}
\begin{abstract}
The Birnbaum--Saunders regression model is commonly used in reliability studies.
We address the issue of performing inference in this class of models when the
number of observations is small. Our simulation results suggest that the likelihood 
ratio test tends to be liberal when the sample size is small. We obtain a correction 
factor which reduces the size distortion of the test. Also, we consider a parametric 
bootstrap scheme to obtain improved critical values and improved $p$-values for the 
likelihood ratio test. The numerical results show that the modified tests are more 
reliable in finite samples than the usual likelihood ratio test. We also present an 
empirical application. 

\end{abstract}
\end{frontmatter}

\section{Introduction}\label{introduction}

Different models have been proposed for lifetime data, such as those based on
the gamma, lognormal and Weibull distributions. These models typically provide a
satisfactory fit in the middle portion of the data, but oftentimes fail to deliver a
good fit at the tails, where only a few observations are generally available.
The family of distributions proposed by Birnbaum and Saunders (1969) can also be
used to model lifetime data. It has the appealing feature of providing satisfactory
tail fitting. This family of distributions was originally obtained from a model
in which failure follows from the development and growth of a dominant crack. 
It was later derived by Desmond (1985) using a biological model which followed
from relaxing some of the assumptions originally made by Birnbaum and Saunders (1969). 

The random variable $T$ is said to be Birnbaum--Saunders distributed with
parameters $\alpha, \eta > 0$, denoted $\mathcal{B}$-$\mathcal{S}(\alpha, \eta)$,
if its distribution function is given by 
\begin{equation}\label{eq11}
F_{T}(t) =\Phi\Biggl[\frac{1}{\alpha}\Biggl(\sqrt{\frac{t}{\eta}}
               - \sqrt{\frac{\eta}{t}}\Biggr)\Biggr],\quad t > 0,
\end{equation}
where $\Phi(\cdot)$ is the standard normal distribution function; 
$\alpha$ and $\eta$ are shape and scale parameters, respectively.
It is easy to show that $\eta$ is the median of the distribution:
$F_{T}(\eta) = \Phi(0) = 1/2$.
For any $k > 0$, it follows that $kT \sim\mathcal{B}$-$\mathcal{S}(\alpha, k\eta)$.
It is also noteworthy that the reciprocal property holds: 
$T^{-1} \sim\mathcal{B}$-$\mathcal{S}(\alpha, \eta^{-1})$,
which is in the same family of distributions [Saunders (1974)].

Rieck and Nedelman (1991) proposed a log-linear regression model based on the
Birnbaum--Saunders distribution. They showed that if $T\sim\mathcal{B}$-$\mathcal{S}(\alpha, \eta)$,
then $y = \log(T)$ is sinh-normal distributed with shape, location and scale parameters
given by $\alpha$, $\mu = \log(\eta)$ and $\sigma = 2$, respectively
[$y\sim\mathcal{SN}(\alpha, \mu, \sigma)$]; see Section~\ref{regBS} for further details. 
Their model has been widely used and is
an alternative to the usual gamma, lognormal and Weibull regression models;
see Rieck and Nedelman (1991, \S~7). Diagnostic tools for the
Birnbaum--Saunders regression model were developed by Galea et al.~(2004), Leiva et al.~(2007)
and Xie and Wei (2007), and Bayesian inference was developed by Tisionas (2001).

Hypothesis testing inference is usually performed using the likelihood
ratio test. It is well known, however, that the limiting null distribution
($\chi^2$) used in the test can be a poor approximation to the exact
null distribution of the test statistic when the number of observations
is small, thus yielding a size-distorted test; see, e.g., the simulation
results in Rieck and Nedelman (1991, \S~5). Consider, for instance,
the application in which interest lies in modeling the die lifetime ($T$) in
the process of metal extrusion, as in Lepadatu et al.~(2005). As noted
by the authors, the die life is mainly determined by its material
properties and the stresses under load. They also note that the extrusion
die is exposed to high temperatures, which can also be damaging. 
The covariates are the friction
coefficient ($x_1$), the angle of the die ($x_2$) and work temperature
($x_3$). Consider a regression model which also includes interaction
effects, i.e.,
\[
y_{i} = \beta_{0} + \beta_{1}x_{1i} + \beta_{2}x_{2i} + \beta_{3}x_{3i} + \beta_{4}x_{1i}x_{2i}
        + \beta_{5}x_{1i}x_{3i} + \beta_{6}x_{2i}x_{3i} + \varepsilon_{i},
\]
where $y_{i} = \log(T_{i})$ and $\varepsilon_{i}\sim\mathcal{SN}(\alpha, 0, 2)$,
$i = 1,2,\ldots,n$. There are only 15 observations ($n=15$), and we wish
to test the significance of the interaction effects, i.e., the interest lies in
testing $\mathcal{H}_{0}\!: \beta_{4} = \beta_{5} = \beta_{6} = 0$. The
likelihood ratio $p$-value equals 0.094, i.e., one rejects the null hypothesis
at the 10\% nominal level. Note, however, that the $p$-value is close to the
significance level of the test and that the number of observations is small.
Can the inference made using the likelihood ratio test be trusted? We
shall return to this application in Section \ref{application}. 

The chief goal of our paper is to improve likelihood ratio inference in
Birnbaum--Saunders regressions when the number of observations available to
the practitioner is small. We do so by following two different approaches.
First, we derive a Bartlett correction factor
that can be applied to the likelihood ratio test statistic. The exact
null distribution of the modified statistic is generally better approximated
by the limiting null distribution used in the test than that of the unmodified
test statistic. Second, we consider a parametric bootstrap resampling 
scheme to obtain improved critical values and improved $p$-values for 
the likelihood ratio test.

The paper unfolds as follows. Section~\ref{regBS} introduces the Birnbaum--Saunders
regression model. In Section~\ref{bartlett}, we derive a Bartlett correction to the
likelihood ratio test statistic; we give a closed-form expression for the
correction factor in matrix form. Special cases are considered in 
Section~\ref{specialcases}. Numerical evidence of the effectiveness of the
finite sample correction we obtain is presented in Section~\ref{simulations}; 
we also evaluate bootstrap-based inference.
Section~\ref{application} addresses the empirical application introduced above
(inferences on die lifetime in metal extrusion). Finally, concluding remarks
are offered in Section~\ref{conclusions}.

\section{The Birnbaum--Saunders regression model}\label{regBS}

The density function of a Birnbaum--Saunders variate $T$ is
\[
f_{T}(t; \alpha, \eta) = \frac{1}{2\alpha\eta\sqrt{2\pi}}\Biggl[\Biggl(\frac{\eta}{t}\Biggr)^{1/2}
            + \Biggl(\frac{\eta}{t}\Biggr)^{3/2}\Biggr]\exp\Biggl\{-\frac{1}{2\alpha^2}
            \Biggl(\frac{t}{\eta} + \frac{\eta}{t} - 2\Biggr)\Biggr\},
\]
where $t,\alpha,\eta > 0$.
The density is right skewed, the skewness decreasing with $\alpha$; see 
Lemonte et al.~(2007, \S~2).
The mean and variance of $T$ are, respectively,
\[
\Es(T) = \eta\biggl(1 + \frac{1}{2}\alpha^2\biggr)\quad{\rm and}\quad{\rm Var}(T) = (\alpha\eta)^2
               \biggl(1 + \frac{5}{4}\alpha^2\biggr).
\]

McCarter~(1999) considered $\mathcal{B}$-$\mathcal{S}(\alpha, \eta)$ parameter
estimation under type II data censoring. Lemonte et al.~(2007) derived the second
order biases of the maximum likelihood estimators of $\alpha$ and $\eta$, and
obtained a corrected likelihood ratio statistic for testing hypotheses regarding
$\alpha$. Lemonte et al.~(2008) proposed several bootstrap bias-corrected
estimators of $\alpha$ and $\eta$. Further details on the Birnbaum--Saunders
distribution can be found in Johnson et al.~(1995).

The $\mathcal{B}$-$\mathcal{S}(\alpha, \eta)$ survival function is
$S_{T}(t) = 1 - F_{T}(t)$, where $F_{T}(t)$ is given in~(\ref{eq11}).
The hazard function is $\nu(t) = {f_{T}(t)}/{S_{T}(t)}$,
where $f_{T}(t)$ is the corresponding density function. The hazard
function $\nu(t)$ equals zero at $t = 0$, increases up to a maximum value and then
decreases towards a given positive level; see Kundu et al.~(2008).
For a comparison between the Birnbaum--Saunders and lognormal hazard functions, see Nelson~(1990).

As noted in the previous section, Rieck and Nedelman (1991) showed that
if $T\sim\mathcal{B}$-$\mathcal{S}(\alpha, \eta)$,
then $y = \log(T)$ follows a sinh-normal distribution with the following
shape, location and scale parameters:
$\alpha$, $\mu = \log(\eta)$ and $\sigma = 2$, respectively, denoted 
$y\sim\mathcal{SN}(\alpha, \mu, \sigma)$. The density function of $y$ is
\[
f(y; \alpha,\mu,\sigma) = \frac{2}{\alpha\sigma\sqrt{2\pi}}\cosh\biggl(\frac{y -
\mu}{\sigma}\biggr)\exp\biggl\{-\frac{2}{\sigma^2}
\mathrm{sinh}^2\biggl(\frac{y - \mu}{\sigma}\biggr)\biggr\}, \quad y\in\mathrm{I\!R}.
\]
This distribution has a number of interesting and attractive properties
[see Rieck (1989)]:
(i) It is symmetric around the location parameter $\mu$; (ii) It is unimodal for 
$\alpha\leq 2$ and bimodal for $\alpha > 2$; (iii) The mean and variance of $y$
are $\Es(y) = \mu$ and Var$(y) = \sigma^2 w(\alpha)$, respectively.
There is no closed-form expression for 
$w(\alpha)$, but Rieck (1989) obtained asymptotic approximations for both small
and large values of $\alpha$;
(iv) If $y_{\alpha}\sim\mathcal{SN}(\alpha, \mu, \sigma)$, then
$S_{\alpha} = 2(y_{\alpha} - \mu)/(\alpha\sigma)$ converges in distribution to the
standard normal distribution when $\alpha\to 0$.

Rieck and Nedelman (1991) proposed the following regression model: 
\begin{equation}\label{eq2:1}
y_{i} = \bsy{x}_{i}^{\top}\bsy{\!\beta} + \varepsilon_{i},\quad i = 1, 2, \ldots, n,
\end{equation}
where $y_{i}$ is the logarithm of the $i$th observed lifetime,
$\bsy{x}_{i}^{\top} = (x_{i1}, x_{i2}, \ldots, x_{ip})$ contains the $i$th
observation on the $p$ covariates ($p < n$),
$\bsy{\beta} = (\beta_1, \beta_2, \ldots, \beta_p)^{\top}$ is a vector of
unknown regression parameters, and $\varepsilon_{i}\sim\mathcal{SN}(\alpha, 0, 2)$.

The log-likelihood function for a random sample $\bsy{y} = (y_1, \ldots, y_n)^{\top}$
from~(\ref{eq2:1}) can be written as 
\begin{equation}\label{eq2:2}
\ell(\bsy{\theta}; \bsy{y}) = -\frac{n}{2}\log(8\pi) + \sum_{i=1}^{n}\log(\xi_{i1}) -
\frac{1}{2}\sum_{i=1}^{n}\xi_{i2}^{2},
\end{equation}
where $\bsy{\theta} = (\bsy{\beta}^{\top}, \alpha)^{\top}$,
\[
\xi_{i1}(\bsy{\theta}) = \xi_{i1} = \frac{2}{\alpha}\cosh\biggl(\frac{y_i - \mu_{i}}{2}\biggr),\quad
\xi_{i2}(\bsy{\theta}) = \xi_{i2} = \frac{2}{\alpha}\mathrm{sinh}\biggl(\frac{y_i - \mu_{i}}{2}\biggr)
\]
and $\mu_{i} = \bsy{x}_{i}^{\top}\bsy{\!\beta}$, $i = 1,2,\ldots, n$.
By differentiating~(\ref{eq2:2}) with respect to $\beta_{r}$ and $\alpha$, we obtain 
\[
\frac{\partial\ell(\bsy{\theta})}{\partial\beta_{r}} =
        \frac{1}{2}\sum_{i=1}^{n}x_{ir}\biggl\{\xi_{i1}\xi_{i2} - \frac{\xi_{i2}}{\xi_{i1}}\biggr\},
\quad r = 1,2,\ldots,p,
\]
and
\[
\frac{\partial\ell(\bsy{\theta})}{\partial\alpha} =
        -\frac{n}{\alpha} + \frac{1}{\alpha}\sum_{i=1}^{n}\xi_{i2}^{2}.
\]
The score function for $\bsy{\beta}$ can be written in matrix form as 
\[
\bsy{U}_{\bsy{\!\beta}}(\bsy{\theta})=\bsy{U}_{\bsy{\!\beta}}
            = \frac{\partial\ell(\bsy{\theta})}{\partial\bsy{\beta}} =  \frac{1}{2}\bsy{X}^{\top}\!\bsy{s},
\]
where $\bsy{X} = (\bsy{x}_1\ \bsy{x}_2\cdots\ \bsy{x}_n)^{\top}$ is the $n\times p$
design matrix (which is assumed to have full column rank) and $\bsy{s}=\bsy{s}(\bsy{\theta})$
is an $n$-vector whose $i$th element equals $\xi_{i1}\xi_{i2} - \xi_{i2}/\xi_{i1}$.

Rieck and Nedelman (1991) obtained a closed-form expression for the maximum
likelihood estimator (MLE) of $\alpha^2$:
\[
\widehat{\alpha}^2 = \frac{4}{n}\sum_{i=1}^{n}\mathrm{sinh}^{2}\biggr(\frac{y_{i} -
                        \bsy{x}_{i}^{\top}\widehat{\bsy{\!\beta}}}{2}\biggl),
\]
where $\widehat{\!\bsy{\beta}}$ is the MLE of $\bsy{\beta}$.
There is no closed-form expression for the MLE of $\bsy{\beta}$.
Hence, one has to use a nonlinear optimization method, such as 
Newton-Raphson or Fisher's scoring, to obtain
$\widehat{\bsy{\!\beta}}$.\footnote{All log-likelihood maximizations with
respect to $\bsy{\beta}$ and $\alpha$ in this paper were carried out using
the BFGS quasi-Newton method with analytic first derivatives; see Press et al.~(1992).
The initial values in the iterative BFGS scheme were 
$\widetilde{\!\bsy{\beta}} = (\bsy{X}^{\top}\!\bsy{X})^{-1}\!\bsy{X}^{\top}\!\bsy{y}$
for $\bsy{\beta}$ and $\sqrt{\widetilde{\alpha}^2}$
for $\alpha$, where $\widetilde{\alpha}^2$ is obtained from $\widehat{\alpha}^2$ with
$\widehat{\!\bsy{\beta}}$ replaced by $\widetilde{\!\bsy{\beta}}$.}

Let $\widehat{\!\bsy{\theta}} = (\,\widehat{\!\bsy{\beta}}{\vspace{-1cm}}^{\,\top},
\widehat{\alpha})^{\top}$ be the MLE of $\bsy{\theta} = (\bsy{\beta}^{\top}, \alpha)^{\top}$.
Rieck and Nedelman (1991) showed that\, $\widehat{\!\bsy{\theta}}\stackrel{A}{\sim}\mathcal{N}_{p+1}(\bsy{\theta},
\bsy{K}(\bsy{\theta})^{-1})$, when $n$ is large, $\stackrel{A}{\sim}$ denoting approximately distributed;
$\bsy{K}(\bsy{\theta})$ is Fisher's information matrix and $\bsy{K}(\bsy{\theta})^{-1}$
is its inverse. Also, $\bsy{K}(\bsy{\theta})$ is a block-diagnonal matrix 
given by $\bsy{K}(\bsy{\theta}) = \mathrm{diag}\{\bsy{K}(\bsy{\beta}),
\kappa_{\alpha, \alpha}\}$: 
$\bsy{K}(\bsy{\beta}) = \psi_{1}(\alpha)(\bsy{X}^{\top}\!\bsy{X})/4$ is Fisher's information for
$\bsy{\beta}$ and $\kappa_{\alpha, \alpha} = 2n/\alpha^2$ is the information relative to
$\alpha$. Also, 
\begin{equation}\label{Fisher}
\psi_{0}(\alpha) = \left\{1 - \mathtt{erf}\left(\frac{\sqrt{2}}{\alpha}\right)\right\} \,
            \exp\left(\frac{2}{\alpha^2}\right)\ {\rm and}\
\psi_{1}(\alpha) = 2 + \frac{4}{\alpha^{2}} - \frac{\sqrt{2\pi}}{\alpha}\psi_{0}(\alpha),
\end{equation}
${\tt erf}(\cdot)$ denoting the error function:
\[
\mathtt{erf}(x) = \frac{2}{\sqrt{\pi}}\int_{0}^{x}\mathrm{e}^{-t^2}\mathrm{d}t.
\]
Details on $\mathtt{erf}(\cdot)$ can be found in Gradshteyn and Ryzhik (2007).
Since $\bsy{K}(\bsy{\theta})$ is block-diagonal, 
$\bsy{\beta}$ and $\alpha$ are globally orthogonal [Cox and Reid (1987)] and
$\widehat{\!\bsy{\beta}}$ and $\widehat{\alpha}$ are asymptotically independent.
It can be shown that when $\alpha$ is small, $\psi_0(\alpha)\approx \alpha/\sqrt{2\pi}$ and
$\psi_1(\alpha)\approx 1 + 4/\alpha^2$; when $\alpha$ is large, $\psi_0(\alpha)\approx 1$
and $\psi_1(\alpha)\approx 2$.

\section{An improved likelihood ratio test}\label{bartlett}

Consider a parametric model $f(\bsy{y}; \bsy{\theta})$ with corresponding
log-likelihood function $\ell(\bsy{\theta}; \bsy{y})$, where $\bsy{\theta} =
(\bsy{\theta}_{1}^{\top}, \bsy{\theta}_{2}^{\top})^{\top}$ is a $k$-vector
of unknown parameters. The dimensions of $\bsy{\theta}_{1}$ and $\bsy{\theta}_{2}$
are $k - q$ and $q$, respectively.
Suppose the interest lies in testing the composite null hypothesis 
${\mathcal H_{0}}\!\!: \bsy{\theta}_{2} = \bsy{\theta}_{2}^{(0)}$
against ${\mathcal H_{2}}\!\!:\bsy{\theta}_{2}\neq\bsy{\theta}_{2}^{(0)}$,
where $\bsy{\theta}_{2}^{(0)}$ is a given vector of scalars. 
Hence, $\bsy{\theta}_{1}$ is a vector of nuisance parameters.
The log-likelihood ratio test statistic can be written as 
\begin{equation}\label{lr1}
LR = 2\bigl\{\ell(\,\widehat{\!\bsy{\theta}}; \bsy{y}) - \ell(\,\widetilde{\!\bsy{\theta}}; \bsy{y})\bigr\},
\end{equation}
where $\widehat{\!\bsy{\theta}} = (\,\widehat{\!\bsy{\theta}}{\vspace{-1cm}}_{1}^{\,\top},
\,\widehat{\!\bsy{\theta}}{\vspace{-1cm}}_{2}^{\,\top})^{\top}$ and
$\widetilde{\!\bsy{\theta}} = (\,\widetilde{\!\bsy{\theta}}{\vspace{-1cm}}_{1}^{\,\top},
\bsy{\theta}_{2}^{(0)\top})^{\top}$ are the MLEs
of $\bsy{\theta} = (\bsy{\theta}_{1}^{\top},
\bsy{\theta}_{2}^{\top})^{\top}$ obtained from the maximization of 
$\ell(\bsy{\theta}; \bsy{y})$ under ${\mathcal H_{1}}$
and ${\mathcal H_{0}}$, respectively.

Bartlett (1937) computed the expected value of $LR$ under ${\mathcal H_{0}}$ up to
order $n^{-1}$: 
$\Es(LR) = q + B(\bsy{\theta}) + O(n^{-2})$,
where $B(\bsy{\theta})$ is a constant of order $O(n^{-1})$.  
It is possible to show that, under the null hypothesis, the mean of
the modified test statistic
\[
LR_{b} = \frac{LR}{1 + B(\bsy{\theta})/q}
\]
equals $q$ when we neglect terms of order $O(n^{-2})$. The order of
the approximation remains unchanged when the unknown parameters
in $B(\bsy{\theta})$ are replaced by their restricted MLEs. Additionally, whereas
$\Pr(LR\leq z) = \Pr(\chi_{q}^{2}\leq z) + O(n^{-1})$, it follows that
$\Pr(LR_b\leq z) = \Pr(\chi_{q}^{2}\leq z) + O(n^{-2})$, a clear improvement. 
The correction factor $c = 1 + B(\bsy{\theta})/q$ is commonly refered to as
the `Bartlett correction factor'. 

Note that $LR$ can be written as 
\[
LR = 2\bigl\{\ell(\,\widehat{\!\bsy{\theta}}; \bsy{y}) - \ell(\bsy{\theta}; \bsy{y})\bigr\} -
     2\bigl\{\ell(\,\widetilde{\!\bsy{\theta}}; \bsy{y})- \ell(\bsy{\theta}; \bsy{y})\bigr\},
\]
where $\ell(\bsy{\theta}; \bsy{y})$ is the log-likelihood function at the
true parameter values. Lawley~(1956) has shown that 
\[
2\, \Es\bigl[\ell(\,\widehat{\!\bsy{\theta}}; \bsy{y}) - \ell(\bsy{\theta}; \bsy{y})\bigr] =
                                            k + \epsilon_{k} + O(n^{-2}),
\]
where $\epsilon_{k}$ is of order $O(n^{-1})$ and is given by 
\begin{equation}\label{lr2}
\epsilon_{k} = \sideset{}{^{\prime}}\sum(\lambda_{rstu} - \lambda_{rstuvw}),
\end{equation}
where $\sum^{\prime}$ denotes summation over all components of $\bsy{\theta}$, i.e.,
the indices $r, s, t, u, v$ and $w$ vary over all $k$ parameters, and the $\lambda$'s
are given by 
\begin{equation}\label{lr3}
\begin{split}
\lambda_{rstu} & = \kappa^{rs}\kappa^{tu}\Bigl\{\frac{\kappa_{rstu}}{4} - \kappa_{rst}^{(u)} + \kappa_{rt}^{(su)}\Bigr\},\\ 
\lambda_{rstuvw} & = \kappa^{rs}\kappa^{tu}\kappa^{vw}\Bigl\{\kappa_{rtv}\Bigl(\frac{\kappa_{suw}}{6} - \kappa_{sw}^{(u)}\Bigr)\\
                 & + \kappa_{rtu}\Bigl(\frac{\kappa_{svw}}{4} - \kappa_{sw}^{(v)}\Bigr) + \kappa_{rt}^{(v)}\kappa_{sw}^{(u)}
                   + \kappa_{rt}^{(u)}\kappa_{sw}^{(v)}\Bigr\}, 
\end{split}
\end{equation}
where $\kappa_{rs} = \Es(\partial^2\ell(\bsy{\theta})/\partial\theta_r\partial\theta_s)$,
$\kappa_{rst} = \Es(\partial^3\ell(\bsy{\theta})/\partial\theta_r\partial\theta_s\partial\theta_t)$,
$\kappa_{rs}^{(t)} = \partial\kappa_{rs}/\partial\theta_{t}$, etc., and $-\kappa^{rs}$ is the
$(r,s)$ element of Fisher's information matrix inverse. Analogously, 
\[
2\, \Es\bigl[\ell(\,\widetilde{\!\bsy{\theta}}; \bsy{y}) - \ell(\bsy{\theta}; \bsy{y})\bigr] =
                                            k - q + \epsilon_{k-q} + O(n^{-2}),
\]
where $\epsilon_{k-q}$ is of order $O(n^{-1})$ and is obtained from~(\ref{lr2}) when
the sum $\sum^{\prime}$ only covers the components of $\bsy{\theta}_{1}$, i.e., the
sum ranges over the $k-q$ nuisance parameters, since $\bsy{\theta}_{2}$ is fixed
under ${\mathcal H}_{0}$.

Under ${\mathcal H_{0}}$, $\Es(LR) = q + \epsilon_{k} - \epsilon_{k-q} + O(n^{-2})$.
Thus, it is possible to achieve a better $\chi_{q}^{2}$ approximation by using
the modified test statistic $LR_b = LR/c$ instead of $LR$, the Bartlett correction factor
being $c = 1 + B(\bsy{\theta})/q$, where $B(\bsy{\theta}) = \epsilon_{k} - \epsilon_{k-q}$.
The corrected statistic $LR_b$ is $\chi_{q}^{2}$ distributed up to order $O(n^{-1})$ under
${\mathcal H}_{0}$. The improved test follows from the comparison of $LR_b$ and
the critical value obtained as the appropriate $\chi_{q}^{2}$ quantile. 

The corrected test statistic is usually written as $LR_{b} = LR/\{1 + B(\bsy{\theta})/q\}$.
Nonetheless, there are alternative modified statistics that are equivalent to $LR_{b}$
to order $O(n^{-1})$, such as $LR_{b}^{*} = LR\exp\{-B(\bsy{\theta})/q\}$ and
$LR_{b}^{**} = LR\{1 - B(\bsy{\theta})/q\}$. It is noteworthy that $LR_{b}^{*}$
has an advantage over the other two specifications: it never assumes negative
values. See Cribari--Neto and Cordeiro (1996) for further details on Bartlett corrections.

In what follows, we shall derive the Bartlett correction factor for testing inference
in the Birnbaum--Saunders regression model. The parameter vector is $\bsy{\theta} =
(\bsy{\beta}^{\top}, \alpha)^{\top}$, which is $(p+1)$-dimensional. Hence,
we shall obtain $\epsilon_{p+1}$ from~(\ref{lr2}), with the indices varying from 1 up to $p+1$.

Let $\bsy{Z} = \bsy{X}(\bsy{X}^{\top}\!\bsy{X})^{-1}\!\bsy{X}^{\top} = \{z_{ij}\}$ and
$\bsy{Z}_{\!d} = \mathrm{diag}\{z_{11}, z_{22}, \ldots, z_{nn}\}$. Also, 
$\bsy{Z}^{(2)} = \bsy{Z}\odot\bsy{Z}$, $\bsy{Z}_{\!d}^{(2)} = \bsy{Z}_{\!d}\odot\bsy{Z}_{\!d}$,
etc., $\odot$ denoting the Hadamard (elementwise) product of matrices.
We shall use the following notation for cumulants of log-likelihood derivatives
with respect to $\bsy{\beta}$ and $\alpha$: $U_{r} = \partial\ell(\bsy{\theta})/\partial\beta_{r}$,
$U_{\alpha} = \partial\ell(\bsy{\theta})/\partial\alpha$,
$U_{rs} = \partial^{2}\ell(\bsy{\theta})/\partial\beta_{r}\partial\beta_{s}$,
$U_{r\alpha} = \partial^{2}\ell(\bsy{\theta})/\partial\beta_{r}\partial\alpha$,
$U_{\alpha\alpha} = \partial^{2}\ell(\bsy{\theta})/\partial\alpha^2$,
$U_{rst} = \partial^{3}\ell(\bsy{\theta})/\partial\beta_{r}\partial\beta_{s}\partial\beta_{t}$, 
$U_{rs\alpha} = \partial^{3}\ell(\bsy{\theta})/\partial\beta_{r}\partial\beta_{s}\partial\alpha$, etc;
$\kappa_{rs} = \Es(U_{rs})$, $\kappa_{r\alpha} = \Es(U_{r\alpha})$, $\kappa_{rst} = \Es(U_{rst})$,
etc; $\kappa_{rs}^{(t)} = \partial\kappa_{rs}/\partial\beta_{t}$,
$\kappa_{r\alpha}^{(t\alpha)} = \partial^{2}\kappa_{r\alpha}/\partial\beta_{t}\partial\alpha$, etc.

From the log-likelihood function in~(\ref{eq2:2}) we obtain the
following cumulants: 
\[
\kappa_{rs} = -\frac{\psi_{1}(\alpha)}{4}\sum_{i=1}^{n}x_{ir}x_{is},\quad
\kappa_{r\alpha} = 0,\quad\kappa_{\alpha\alpha} = -\frac{2n}{\alpha^2},
\]
\[
\kappa_{rst} = 0,\quad
\kappa_{rs\alpha} = \frac{2+\alpha^2}{\alpha^3}\sum_{i=1}^{n}x_{ir}x_{is},\quad
\kappa_{r\alpha\alpha} = 0,\quad
\kappa_{\alpha\alpha\alpha} = \frac{10n}{\alpha^3},
\]
\[
\kappa_{rstu} = \psi_{2}(\alpha)\sum_{i=1}^{n}x_{ir}x_{is}x_{it}x_{iu},\quad
\kappa_{rst\alpha} = 0,\quad
\kappa_{rs\alpha\alpha} = -\frac{3(2 + \alpha^2)}{\alpha^4}\sum_{i=1}^{n}x_{ir}x_{is},
\]
\[
\kappa_{r\alpha\alpha\alpha} = 0\quad{\rm and}\quad
\kappa_{\alpha\alpha\alpha\alpha} = -\frac{54n}{\alpha^4},
\]
where 
\[
\psi_{2}(\alpha) = -\frac{1}{4}\biggl\{2 + \frac{7}{\alpha^2} -\sqrt{\frac{\pi}{2}}\biggl(\frac{1}{2\alpha}
+ \frac{6}{\alpha^3}\biggr)\psi_{0}(\alpha)\biggr\}
\]
and $\psi_{0}(\alpha)$ and $\psi_{1}(\alpha)$ are defined in~(\ref{Fisher}).
For small $\alpha$, we have $\psi_2(\alpha)\approx -5/8 - 1/\alpha^2$;
for large $\alpha$, $\psi_2(\alpha)\approx -1/2$.

Using these cumulants and also making use of the orthogonality between
$\bsy{\beta}$ and $\alpha$, we obtain, after long and tedious algebra (Appendix),
$\epsilon_{p+1} = \epsilon(\alpha, p, \bsy{X})$, where
\begin{equation}\label{bartlett:geral}
\epsilon(\alpha, p, \bsy{X}) = \epsilon_{\alpha}(\alpha, p) + \epsilon_{\bsy{\beta}}(\alpha, \bsy{X}),
\end{equation}
with 
\[
\epsilon_{\alpha}(\alpha, p) = \frac{1}{n}\biggl\{\frac{1}{3} + \delta_{1}(\alpha)p
+ \delta_{2}(\alpha)p^2\biggr\}\quad{\rm and}\quad
\epsilon_{\bsy{\beta}}(\alpha, \bsy{X}) = \delta_{3}(\alpha)\mathrm{tr}(\bsy{Z}_{\!d}^{(2)}).
\]
Here, $\mathrm{tr}(\cdot)$ denotes the trace operator and
\[
\delta_{0}(\alpha) = \frac{2 + \alpha^2}{\psi_{1}(\alpha)\alpha^2}, \quad
\delta_{1}(\alpha) = 4\delta_{0}(\alpha)\biggl\{\frac{2}{2 + \alpha^2}
+ \delta_{0}(\alpha) - \frac{2\alpha\psi_{3}(\alpha)}{\psi_{1}(\alpha)}\biggr\},
\]
\[
\delta_{2}(\alpha) = 2\delta_{0}(\alpha)^2,\
\delta_{3}(\alpha) = \frac{4\psi_{2}(\alpha)}{\psi_{1}(\alpha)^2}\ {\rm and}\
\psi_{3}(\alpha) = \frac{3}{\alpha^3} - \frac{\sqrt{2\pi}}{4\alpha^2}\biggr(1 + \frac{4}{\alpha^2}\biggl)\psi_{0}(\alpha).
\]

In expression~(\ref{bartlett:geral}) -- our main result -- we write
$\epsilon_{p+1}$ as the sum of two terms,
namely $\epsilon_{\alpha}(\alpha, p)$ and $\epsilon_{\bsy{\beta}}(\alpha, \bsy{X})$.
The quantity $\epsilon_{\bsy{\beta}}(\alpha, \bsy{X})$ is obtained from (6) with
$\sum'$ ranging over the components of $\bsy{\beta}$, i.e. as if $\alpha$ were known.
The quantity $\epsilon_{\alpha}(\alpha, p)$ is the contribution yielded by the
fact that $\alpha$ is unknown (see the Appendix). Note that $\epsilon_{\alpha}(\alpha, p)$
depends on the design matrix only through its rank. More specifically, it is
a second degree polynomial in $p$ divided by $n$.
Hence, $\epsilon_{\alpha}(\alpha, p)$ can be non-negligible if the dimension of
$\bsy{\beta}$ is not considerably smaller than the sample size.
It is also noteworthy that $\epsilon(\alpha, p, \bsy{X})$ depends on $\alpha$ but not on
$\bsy{\beta}$. The dependency of $\epsilon(\alpha, p, \bsy{X})$ on $\alpha$ occurs through
$\delta_1(\alpha)$, $\delta_2(\alpha)$ and $\delta_3(\alpha)$.
For small $\alpha$, we have
$\delta_1(\alpha) \approx 1$,
$\delta_2(\alpha)\approx 1/2$
and
$\delta_3(\alpha)\approx 0$.
For large $\alpha$,  
$\delta_1(\alpha) \approx 1$, $\delta_2(\alpha)\approx 1/2$ and 
$\delta_3(\alpha)\approx -1/2$. 
Furthermore, $\mathrm{tr}(\bsy{Z}_{\!d}^{(2)})$ establishes the dependency of 
$\epsilon(\alpha, p, \bsy{X})$ on $\bsy{X}$. In other words, $\epsilon(\alpha, p, \bsy{X})$ 
depends on the sum of squares of the diagonal elements of the hat matrix $\bsy{Z}$. In particular, 
if $p=1$, i.e. if $\bsy{X}$ has a single column, $\bsy{x}=(x_1,\ldots,x_n)^\top$ say, 
then $\mathrm{tr}(\bsy{Z}_{\!d}^{(2)}) = \sum_{i=1}^{n}x_{i}^4/
\bigr(\sum_{i=1}^{n}x_{i}^{2}\bigl)^{2}$, the sample kurtosis of $\bsy{x}$.
                  
Finally, it should be noted that expression~(\ref{bartlett:geral}) is quite
simple and can be easily implemented into any mathematical or statistical/econometric programming
environment, such as {\tt R} [R Development Core Team (2006)], {\tt Ox}
[Cribari--Neto and Zarkos (2003); Doornik (2006)] and {\tt MAPLE} [Abell and Braselton (1994)].

\section{Special cases}\label{specialcases}

In this section we present closed-form expressions for the Bartlett correction
factor in situations that are of particular interest to practitioners. The
simplified expressions are obtained from our more general result given
in~(\ref{bartlett:geral}).

At the outset, we consider the test of 
$\mathcal{H}_{0}\!\!:\alpha = \alpha^{(0)}$ against
$\mathcal{H}_{1}\!\!:\alpha\neq\alpha^{(0)},$
where $\alpha^{(0)}$ is a given positive scalar and 
$\bsy{\beta}$ is a vector of nuisance parameters. The Bartlett correction factor
becomes $c = 1 + B(\bsy{\theta})$, where
$B(\bsy{\theta}) = \epsilon(\alpha, p, \bsy{X}) - \epsilon_{\bsy{\beta}}(\alpha, \bsy{X})$,
and hence, $B(\bsy{\theta}) = \epsilon_{\alpha}(\alpha, p)$. 
Note that the correction factor depends on $\bsy{X}$ only through its rank, $p$.
In particular, when $p=1$ (i.i.d.~case), we have
\[
B(\bsy{\theta}) = \frac{1}{n}\biggl\{\frac{1}{3} + \delta_{1}(\alpha)
+ \delta_{2}(\alpha)\biggr\}.
\]
This formula corrects eq.~(14) in Lemonte et al.~(2007), which is in error.
For small and large values of $\alpha$, we have $B(\bsy{\theta}) \approx 11/(6n)$.

Oftentimes practitioners wish to test restrictions on a subset of the
regression parameters. For instance, one may want to test whether a
given group of covariates are jointly significant. To that end, we
partition $\bsy{\beta}$ as 
$\bsy{\beta} = (\bsy{\beta}_{1}^{\top}, \bsy{\beta}_{2}^{\top})^{\top}$, where
$\bsy{\beta}_{1} = (\beta_{1}, \beta_{2},\dots,\beta_{p-q})^{\top}$ and 
$\bsy{\beta}_{2} = (\beta_{p-q+1}, \beta_{p-q+2},\dots,\beta_{p})^{\top}$ are vectors
of dimensions $(p - q)\times 1$ and $q\times 1$, respectively, and consider the test of
$\mathcal{H}_{0}\!\!:\bsy{\beta}_{2} = \bsy{\beta}_{2}^{(0)}$ against
$\mathcal{H}_{1}\!\!:\bsy{\beta}_{2}\neq\bsy{\beta}_{2}^{(0)}$,
where $\bsy{\beta}_{2}^{(0)}$ is a $q$-vector of known constants. The most common
situation is that in which $\bsy{\beta}_{2}^{(0)} = {\bf 0}$. 
Note that $\bsy{\beta}_{1}$ and $\alpha$ are nuisance parameters.
In accordance with the partition of $\bsy{\beta}$, we partition $\bsy{X}$ as
$\bsy{X} = (\bsy{X}_{\!1}\ \bsy{X}_{\!2})$, where
the dimensions of $\bsy{X}_{\!1}$ and $\bsy{X}_{\!2}$ are $n\times(p-q)$ and $n\times q$,
respectively. The correction factor is $c = 1 + B(\bsy{\theta})/q$, where
$B(\bsy{\theta}) = \epsilon(\alpha, p, \bsy{X}) - \epsilon(\alpha, p-q, \bsy{X}_{\!1})$.
It is easy to obtain
\[
B(\bsy{\theta}) = \frac{1}{n}\bigl\{\delta_1(\alpha) q + \delta_2(\alpha) q(2p-q)\bigr\} 
                 + \delta_{3}(\alpha)\mathrm{tr}(\bsy{Z}_{\!d}^{(2)} - \bsy{Z}_{1\!d}^{(2)}),
\]
with
$\bsy{Z}_{\!1} = \bsy{X}_{\!1}(\bsy{X}_{\!1}^{\top}\!\bsy{X}_{\!1})^{-1}\!\bsy{X}_{\!1}^{\top} = \{z_{1ij}\}$
and $\bsy{Z}_{\!1d} = {\rm diag}\{z_{111}, z_{122}, \ldots, z_{1nn}\}$.

Next, suppose we wish to test
$\mathcal{H}_{0}\!\!:\bsy{\beta} = \bsy{\beta}^{(0)}$ against
$\mathcal{H}_{1}\!\!:\bsy{\beta}\neq\bsy{\beta}^{(0)}$,
where $\bsy{\beta}^{(0)}$ is a $p$-vector of known constants and $\alpha$ is a nuisance
parameter. The Bartlett correction factor is $c = 1 + B(\bsy{\theta})/p$
with
$B(\bsy{\theta}) = \epsilon(\alpha, p, \bsy{X}) - \epsilon_{\alpha}(\alpha, 0)$,
which yields
\[
B(\bsy{\theta}) = \frac{1}{n}\bigl\{\delta_1(\alpha) p + \delta_2(\alpha) p^2\bigr\} 
                 + \delta_{3}(\alpha)\mathrm{tr}(\bsy{Z}_{\!d}^{(2)}).
\]

\section{Numerical evidence}\label{simulations}

We shall now report Monte Carlo evidence on the finite sample performance of
three tests in Birnbaum--Saunders regressions, namely: the likelihood ratio
test ($LR$), the Bartlett-corrected likelihood ratio test ($LR_{b}$), and an
asymptotically equivalent corrected test ($LR_{b}^{*}$).\footnote{We do not report
results relative to $LR_{b}^{**}$ since they were very similar to those obtained
using $LR_{b}^{*}$.} The model used in the numerical evaluation is 
\[
y_{i} = \beta_{1}x_{i1} + \beta_{2}x_{i2} + \cdots + \beta_{p}x_{ip} + \varepsilon_{i},
\]
where $x_{i1} = 1$ and $\varepsilon_{i}\sim\mathcal{SN}(\alpha, 0, 2)$, $i = 1, 2, \ldots, n$.
The covariate values were selected as random draws from the $\mathcal{U}(0,1)$
distribution. The number of Monte Carlo replications was 10,000, the nominal levels
of the tests were $\gamma$ = 10\%, 5\% and 1\%, and all simulations were carried
out using the \texttt{Ox} matrix programming language (Doornik, 2006).  

Table~\ref{tab1:2} presents the null rejection rates (entries are percentages) of the
three tests. The null hypothesis is $\mathcal{H}_{0}\!\!:\beta_{p-1} =\beta_{p} = 0$,
which is tested against a two-sided alternative, the sample size is $n=30$ and
$\alpha = 0.5$. Different values of $p$ were considered. The values of the response were
generated using $\beta_{1} = \beta_{2} =\cdots=\beta_{p-2} = 1$.

Note that the likelihood ratio test is considerably oversized (liberal), more so
as the number of regressors increases. For instance, when $p = 8$ and $\gamma = 10\%$,
its null rejection rate is 18.78\%, i.e., nearly twice the nominal level of the test.
The two corrected tests are much less size distorted. For example, their null
rejection rates in the same situation were 11.82\% ($LR_{b}$) and 11.13\% ($LR_{b}^{*}$).

\begin{table}[htp]
\renewcommand{\arraystretch}{1.1}                       
\caption{Null rejection rates; $\alpha = 0.5$, $n = 30$.}\label{tab1:2}
\begin{tabular}{c | c c c | c c c | c c c }\hline                                                              
  & \multicolumn{3}{c|}{$\gamma = 10\%$} & \multicolumn{3}{c|}{$\gamma = 5\%$}
  & \multicolumn{3}{c}{$\gamma = 1\%$}  \\\cline{2-10}                              
  $p$   & $LR$  & $LR_{b}$  & $LR_{b}^{*}$ & $LR$  & $LR_{b}$  & $LR_{b}^{*}$                               
        & $LR$  & $LR_{b}$  & $LR_{b}^{*}$\\\hline                  
    3   & 12.69 & 10.36 & 10.22  &  6.51 & 4.98 & 4.90 & 1.75 & 1.25 & 1.23  \\                    
    4   & 13.44 & 10.27 & 10.07  &  7.46 & 5.41 & 5.32 & 1.90 & 1.10 & 1.04  \\                    
    5   & 14.77 & 10.74 & 10.45  &  8.25 & 5.53 & 5.31 & 2.21 & 1.18 & 1.14  \\             
    6   & 15.94 & 11.07 & 10.53  &  9.14 & 5.55 & 5.23 & 2.54 & 1.24 & 1.17  \\                    
    7   & 17.28 & 11.55 & 10.88  & 10.13 & 5.69 & 5.42 & 2.95 & 1.29 & 1.19  \\                    
    8   & 18.78 & 11.82 & 11.13  & 11.15 & 6.44 & 5.83 & 3.38 & 1.48 & 1.31  \\                    
    9   & 19.92 & 12.11 & 11.00  & 12.00 & 6.33 & 5.66 & 3.82 & 1.49 & 1.25  \\\hline                   
\end{tabular}                                                                                                                                                                                                                                                      
\end{table}

The results in Table~\ref{tab3:2} correspond to $\alpha = 0.5$ and $p=6$. We report results
for samples sizes ranging from 20 to 200. The null hypothesis under test is
$\mathcal{H}_{0}\!\!:\beta_{5} =\beta_{6} = 0$. The figures in this table show that the
null rejection rates of all tests approach the corresponding nominal levels as the
sample size grows, as expected. It is also noteworthy that the likelihood ratio
test displays liberal behavior even when $n=100$. Overall, the corrected tests
are less size distorted than the unmodified test. For example, when $n = 50$ and $\gamma = 5\%$,
the null rejection rates are 7.49\% ($LR$), 5.32\% ($LR_{b}$) and 5.17\% ($LR_{b}^{*}$). 

\begin{table}[htp]
\renewcommand{\arraystretch}{1.1}                      
\caption{Null rejection rates; $\alpha = 0.5$,
        $p = 6$ and different sample sizes.}\label{tab3:2}
\begin{tabular}{c | c c c | c c c | c c c}\hline                                                              
                                                                                                          
  & \multicolumn{3}{c|}{$\gamma = 10\%$} & \multicolumn{3}{c|}{$\gamma = 5\%$}
  & \multicolumn{3}{c}{$\gamma = 1\%$} \\\cline{2-10}                              
  $n$   & $LR$  & $LR_{b}$  & $LR_{b}^{*}$ & $LR$  & $LR_{b}$  & $LR_{b}^{*}$                             
        & $LR$  & $LR_{b}$  & $LR_{b}^{*}$\\ \hline                  
    20  & 19.54 & 12.04 & 11.08 & 11.97 & 6.54 & 5.87 & 4.05 & 1.58 & 1.38 \\                    
    30  & 15.94 & 11.07 & 10.53 &  9.14 & 5.55 & 5.23 & 2.54 & 1.24 & 1.17 \\                    
    40  & 13.57 & 10.14 &  9.97 &  7.45 & 4.99 & 4.81 & 1.79 & 1.03 & 1.01 \\                                              
    50  & 13.36 & 10.72 & 10.51 &  7.49 & 5.32 & 5.17 & 1.51 & 1.02 & 0.99 \\                                                                      
   100  & 11.86 & 10.46 & 10.44 &  5.90 & 4.92 & 4.88 & 1.25 & 1.04 & 1.03 \\                                                                                      
   200  & 10.92 & 10.14 & 10.12 &  5.57 & 5.07 & 5.07 & 1.04 & 0.96 & 0.96 \\\hline                                                                   
\end{tabular}                                                                                                                                                                                                                                                                                                                                                        
\end{table}

\begin{figure}
\centering
\includegraphics[width=12cm, height=9cm]{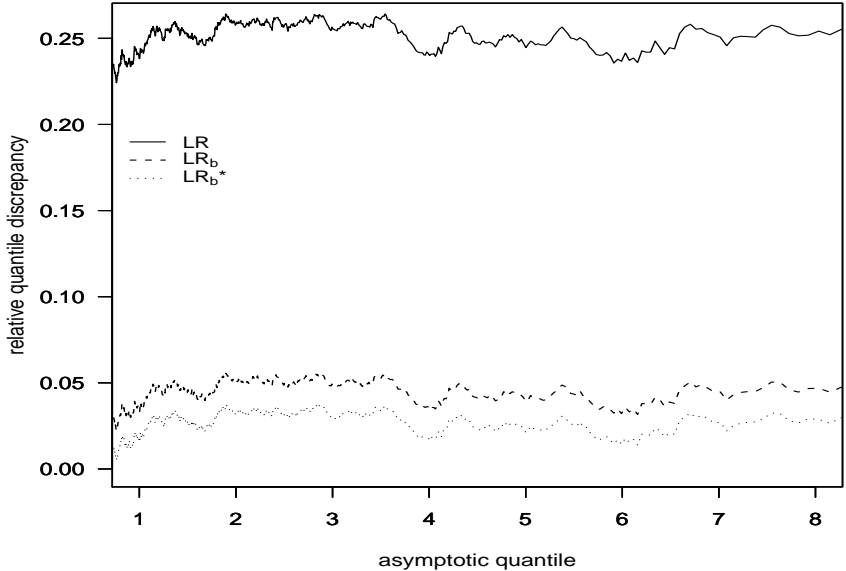}
\caption{Relative quantile discrepancies plot: $n = 30$, $p = 6$ and $\alpha = 0.5$.}\label{figDISC}
\end{figure}

Figure~\ref{figDISC} plots relative quantile discrepancies against the associated
asymptotic quantiles for the three test statistics. Relative quantile discrepancies
are defined as the difference between exact (estimated by Monte Carlo) and asymptotic
quantiles divided by the latter. Again, $p=6$ and we test the exclusion of the last
two covariates. Also, $n=30$ and $\alpha = 0.5$. The closer to zero the relative
quantile discrepancies, the more accurate the test. While Tables 1 and 2 give rejection 
rates of the tests at fixed nominal levels, Figure~\ref{figDISC} compares the
whole distributions of the different statistics with the limiting null distribution.
We note that the relative quantile discrepancies of the likelihood ratio test statistic oscillates around
25\% whereas for the two corrected statistics they are around 5\% ($LR_{b}$) and 3\% ($LR_{b}^*$).
It is thus clear that the null distributions of the modified statistics are much better approximated
by the limiting null distribution ($\chi_2^2$) than that of the likelihood ratio
statistic. 

Table~\ref{tab4:2} contains the nonnull rejection rates (powers) of the tests.  
Here, $p = 4$, $\alpha = 0.5$ and $n = 30, 50, 100$. Data generation was performed
under the alternative hypothesis: $\beta_{3} = \beta_{4} = \delta$, with different
values of $\delta$ ($\delta > 0$). We have only considered the two corrected
tests since the likelihood ratio is considerably oversized, as noted earlier. 
Note that the two tests display similar powers. For instance, when $n = 50$, $\gamma = 5\%$
and $\delta = 0.5$, the nonnull rejection rates are 72.39\% ($LR_{b}$) and 72.28\%
($LR_{b}^*$). We also note that the powers of the tests increase with $n$ and also
with $\delta$, as expected. 

\begin{table}[htp]
\renewcommand{\arraystretch}{1.1}
\caption{Nonnull rejection rates; $\alpha = 0.5$,
        $p = 4$ and different sample sizes.}\label{tab4:2}
\begin{tabular}{c c| c c c| c c c}\hline                                                              
&  & \multicolumn{3}{c|}{$LR_{b}$} & \multicolumn{3}{c}{$LR_{b}^{*}$} \\\cline{3-8}                              
  $n$ & $\delta$ & 10\%  & 5\% & 1\% & 10\%  & 5\% & 1\% \\ \hline
   30 & 0.1 & 13.20 &  6.91 &  1.57 & 13.01 &  6.73 &  1.52  \\            
      & 0.2 & 20.66 & 12.22 &  3.46 & 20.40 & 12.02 &  3.30  \\        
      & 0.3 & 33.07 & 21.63 &  7.49 & 32.73 & 21.28 &  7.33  \\       
      & 0.4 & 48.36 & 35.57 & 14.96 & 48.08 & 35.20 & 14.61  \\            
      & 0.5 & 65.11 & 51.59 & 26.42 & 64.72 & 51.19 & 25.99  \\\hline           
                                                     
   50 & 0.1 & 13.82 &  7.63 &  2.03 & 13.71 &  7.60 &  1.99  \\                  
      & 0.2 & 25.89 & 16.03 &  5.11 & 25.81 & 15.96 &  5.03  \\             
      & 0.3 & 45.00 & 32.06 & 13.15 & 44.86 & 31.95 & 13.07  \\             
      & 0.4 & 65.07 & 52.09 & 28.18 & 64.97 & 51.96 & 28.03  \\               
      & 0.5 & 82.31 & 72.39 & 48.01 & 82.14 & 72.28 & 47.88  \\\hline              
                                                     
  100 & 0.1 & 18.66 & 11.02 &  2.91 & 18.65 & 11.01 &  2.90  \\             
      & 0.2 & 43.63 & 31.29 & 13.05 & 43.61 & 31.28 & 13.02  \\             
      & 0.3 & 73.47 & 62.39 & 37.40 & 73.47 & 62.35 & 37.34  \\         
      & 0.4 & 92.12 & 86.39 & 69.15 & 92.11 & 86.37 & 69.11  \\             
      & 0.5 & 98.76 & 97.34 & 89.98 & 98.76 & 97.33 & 89.93  \\\hline             
\end{tabular}                                                                                                                                                                                                                                                                                                                                          
\end{table}

Table~\ref{tab5:2} presents the null rejection rates for inference on
the scalar parameter $\alpha$.
Here, $n = 30$ and $p$ = 2, 3 and 4. The null hypotheses under test are   
$\mathcal{H}_{0}\!\!: \alpha = 0.5$ and $\mathcal{H}_{0}\!\!: \alpha = 1.0$.
The likelihood ratio test is again liberal.
Note that the two corrected tests are much less size
distorted. For instance, when $p = 4$, $\gamma = 5\%$ and $\alpha = 1.0$,
the null rejection rates of the $LR$, $LR_{b}$ and $LR_{b}^{*}$ tests were
12.03\%, 5.20\% and 4.02\%, respectively. 

\begin{table}[htp]                                                                                            
\renewcommand{\arraystretch}{1.03}
\caption{Null rejection rates; inference on $\alpha$; 
$n = 30$ and different values for $p$.}\label{tab5:2}
\begin{tabular}{c c| c c c | c c c }\hline                                                              
 &  & \multicolumn{3}{c|}{$\mathcal{H}_{0}\!\!: \alpha = 0.5$} &
      \multicolumn{3}{c}{$\mathcal{H}_{0}\!\!: \alpha = 1.0$} \\ \cline{3-8}                              
$p$  &     & 10\% & 5\% & 1\% & 10\% & 5\% & 1\% \\ \hline                  
  2  &  $LR$         & 12.76 &  6.99 & 1.82 & 13.55 &  7.26 & 1.93\\            
     &  $LR_{b}$     & 10.13 &  5.15 & 1.21 & 10.52 &  5.10 & 1.08\\            
     &  $LR_{b}^{*}$ &  9.90 &  5.02 & 1.20 & 10.31 &  4.94 & 1.05\\ \hline     
                                                             
  3  &  $LR$         & 15.16 &  8.64 & 2.45 & 16.10 &  9.35 & 2.70\\            
     &  $LR_{b}$     & 10.46 &  5.43 & 1.16 & 10.53 &  5.06 & 0.97\\            
     &  $LR_{b}^{*}$ &  9.77 &  5.02 & 1.02 &  9.65 &  4.68 & 0.81\\ \hline     
                                                             
  4  &  $LR$         & 18.16 & 10.72 & 3.39 & 19.86 & 12.03 & 3.73\\            
     &  $LR_{b}$     & 10.45 &  5.37 & 0.98 & 10.73 &  5.20 & 0.75\\            
     &  $LR_{b}^{*}$ &  9.29 &  4.57 & 0.72 &  8.77 &  4.02 & 0.43\\ \hline     
\end{tabular}
\vskip 0.2truein
\end{table}

Our simulation results concerning tests on the regression parameters were obtained
for $\alpha=0.5$.  In practice, values of $\alpha$ between 0 and 1 cover most of
the applications; see, for instance, Rieck and Nedelman~(1991). We shall now present
simulation results for a wide range of values of $\alpha$, namely
$\alpha = 0.1, 0.3, 0.5, 0.7,$ $0.9, 1.2, 2, 10, 50$ and $100$.
The new set of simulation results includes rejection rates of the likelihood ratio
test that uses parametric bootstrap critical values (with 600 bootstrap
replications).
The parametric bootstrap can be briefly described as follows.
We can use bootstrap resampling to estimate the null distribution of the 
statistic $LR$ directly from the observed sample 
$\bsy{y} = (y_1,\ldots, y_n)^{\top}$. To that end, one generates, under 
$\mathcal{H}_{0}$ (i.e., imposing the restrictions stated in the 
null hypothesis), $B$ bootstrap samples
$(\bsy{y}^{*1}, \ldots, \bsy{y}^{*B})$ from the assumed model with 
the parameters replaced by restricted estimates computed using 
the original sample (parametric bootstrap), 
and, for each pseudo-sample, one computes 
$LR^{*b} = 2\{\ell(\widehat{\theta}^{*b}; \bsy{y}^{*b}) - \ell(\widetilde{\theta}^{*b}; \bsy{y}^{*b})\}$,
$b = 1, 2, \ldots, B$, where $\widetilde{\theta}^{*b}$ and $\widehat{\theta}^{*b}$ are 
the maximum likelihood estimators of $\theta$ obtained 
from the maximizations of $\ell(\theta; \bsy{y}^{*b})$ under
$\mathcal{H}_{0}$ and $\mathcal{H}_{1}$, respectively.
The $1- \gamma$ percentile of $LR^{*b}$ is estimated by 
$\widehat{q}_{1-\gamma}$, such that 
$\#\{LR^{*b} \leq  \widehat{q}_{1-\gamma}\}/B = 1 - \gamma$.
One rejects the null hypothesis 
if $LR > \widehat{q}_{1-\gamma}$. For a good discussion of 
bootstrap tests, see Efron and Tibshirani (1993, Chapter 16).

Figures in Table~\ref{tab6:2} provide important information. For all values of $\alpha$ the Bartlett
and the bootstrap corrections are very effective in pushing the rejection
rates toward the nominal levels. An advantage of the Bartlett correction over the bootstrap approach is
that the first requires much less computational effort.
It is noteworthy that as $\alpha$ grows, the rejection rates of the likelihood ratio test approaches
the corresponding nominal levels, making the corrections less needed.

\begin{table}[!htp]
\renewcommand{\arraystretch}{1.12}
\caption{Null rejection rates of $\mathcal{H}_{0}\!: \beta_{3} = \beta_{4} = 0$; $p=4$, 
$n = 25$ and different values for $\alpha$.}\label{tab6:2}
\begin{tabular}{c  c c c c| c c c c }\hline
&\multicolumn{4}{c}{$\alpha$ = 0.1}&\multicolumn{4}{c}{$\alpha$ = 0.3}\\\cline{2-9}
  $\gamma$   & $LR$ & $LR_{b}$ & $LR_{b}^{*}$ & $LR_{boot}$
             & $LR$ & $LR_{b}$ & $LR_{b}^{*}$ & $LR_{boot}$ \\\hline
    10\% & 14.33 & 11.42 & 11.32 & 9.90  & 14.45 & 10.70 & 10.44 & 10.18  \\                    
     5\% &  7.88 &  5.82 &  5.74 & 4.94  &  8.01 &  5.49 &  5.22 &  4.97  \\                                     
     1\% &  1.95 &  1.29 &  1.23 & 1.24  &  2.07 &  1.20 &  1.13 &  1.13  \\\hline
&\multicolumn{4}{c}{$\alpha$ = 0.5}&\multicolumn{4}{c}{$\alpha$ = 0.7}\\\cline{2-9}
  $\gamma$   & $LR$ & $LR_{b}$ & $LR_{b}^{*}$ & $LR_{boot}$
             & $LR$ & $LR_{b}$ & $LR_{b}^{*}$ & $LR_{boot}$ \\\hline
    10\% & 14.25 & 10.23 & 10.01 & 10.23 & 14.17 & 10.61 & 10.36 & 10.14  \\                    
     5\% &  7.69 &  5.17 &  5.02 &  5.12 &  8.09 &  5.35 &  5.17 &  5.28  \\                                                            
     1\% &  1.89 &  0.91 &  0.85 &  1.24 &  2.07 &  1.12 &  1.06 &  1.02  \\\hline
&\multicolumn{4}{c}{$\alpha$ = 0.9}&\multicolumn{4}{c}{$\alpha$ = 1.2}\\\cline{2-9}
  $\gamma$   & $LR$ & $LR_{b}$ & $LR_{b}^{*}$ & $LR_{boot}$
             & $LR$ & $LR_{b}$ & $LR_{b}^{*}$ & $LR_{boot}$ \\\hline
    10\% & 13.96 & 10.80 & 10.60 & 9.64  & 13.49 & 10.45 & 10.29 & 9.79 \\                                            
     5\% &  8.03 &  5.77 &  5.61 & 5.17  &  7.51 &  5.37 &  5.26 & 5.10 \\                                     
     1\% &  2.29 &  1.30 &  1.26 & 1.10  &  1.90 &  1.18 &  1.16 & 1.38 \\\hline
&\multicolumn{4}{c}{$\alpha$ = 2}&\multicolumn{4}{c}{$\alpha$ = 10}\\\cline{2-9}
  $\gamma$   & $LR$ & $LR_{b}$ & $LR_{b}^{*}$ & $LR_{boot}$
             & $LR$ & $LR_{b}$ & $LR_{b}^{*}$ & $LR_{boot}$ \\\hline
    10\% & 13.21 & 10.87 & 10.64 & 10.01 & 12.44 & 11.23 & 11.13 & 9.75  \\                    
     5\% &  7.29 &  5.61 &  5.50 &  5.08 &  6.59 &  5.81 &  5.73 & 4.80  \\                                                            
     1\% &  1.63 &  1.08 &  1.04 &  1.21 &  1.42 &  1.17 &  1.16 & 0.98  \\\hline
&\multicolumn{4}{c}{$\alpha$ = 50}&\multicolumn{4}{c}{$\alpha$ = 100}\\\cline{2-9}
  $\gamma$   & $LR$ & $LR_{b}$ & $LR_{b}^{*}$ & $LR_{boot}$
             & $LR$ & $LR_{b}$ & $LR_{b}^{*}$ & $LR_{boot}$ \\\hline
    10\% & 11.26 & 10.45 & 10.43 & 10.11 & 10.87 & 10.17 & 10.12 & 9.89 \\                                            
     5\% &  5.60 &  5.21 &  5.20 &  4.93 &  5.67 &  5.11 &  5.07 & 5.18 \\                                     
     1\% &  1.19 &  1.06 &  1.06 &  1.04 &  1.23 &  1.07 &  1.07 & 1.18 \\\hline
\end{tabular}
\end{table}

We shall now try to shed some light on the issue of the possible effect of near-collinearity  
between the covariates $\bsy{X}$ on the testing procedures.
To do so, we performed an additional simulation experiment. We set $p=4$ and selected the
covariate values as follows: $x_{i1} =1,$ for
$i=1,\ldots,n$, the values of $x_{2}$ were chosen as random draws from the
${\mathcal U}(0,1)$
distribution and the pairs $(x_{i3},x_{i4})$ were selected as random draws from
the bivariate normal distribution $\mathcal{N}_{2}(\bsy{0}, \bsy{\Sigma})$,
where the covariance matriz $\bsy{\Sigma}$ has the following form
\[
\bsy{\Sigma} =
\begin{pmatrix}
 1     & \rho \\
\rho & 1      \\
\end{pmatrix}.
\]
The closer the value of $\rho$ is to either extreme ($-1$ or 1), the stronger the linear relation
between the covariates $x_{3}$ and $x_{4}$. Table \ref{tab7:2} presents simulation results for different values of $\rho$.
The figures in this table suggest that the sample correlation between  $x_2$ and $x_3$ does not
have significant effect on the behaviour of the testing procedures. Hence, near-collinearity
does not seem to a matter of concern. 

\begin{table}[!htp]
\renewcommand{\arraystretch}{1.12}
\caption{Null rejection rates of $\mathcal{H}_{0}\!: \beta_{2} = \beta_{4} = 0$; $p=4$, 
$n = 20$ and different values for $\rho$.}\label{tab7:2}
\begin{tabular}{c  c c c c}\hline
&\multicolumn{4}{c}{$\rho$ = 0.0}\\\cline{2-5}
  $\gamma$   & $LR$ & $LR_{b}$ & $LR_{b}^{*}$ & $LR_{boot}$ \\\hline
    10\% & 16.00 & 11.15 & 10.72 & 10.20 \\                    
     5\% &  9.16 &  5.33 &  5.08 &  4.90 \\                                     
     1\% &  2.37 &  1.27 &  1.21 &  1.16 \\\hline
&\multicolumn{4}{c}{$\rho$ = 0.5}\\\cline{2-5}
  $\gamma$   & $LR$ & $LR_{b}$ & $LR_{b}^{*}$ & $LR_{boot}$ \\\hline
    10\% & 15.54 & 10.72 & 10.33 & 10.14  \\                    
     5\% &  8.85 &  5.73 &  5.48 &  5.22  \\                                     
     1\% &  2.37 &  1.15 &  1.03 &  1.10  \\\hline
&\multicolumn{4}{c}{$\rho$ = 0.9}\\\cline{2-5}
  $\gamma$   & $LR$ & $LR_{b}$ & $LR_{b}^{*}$ & $LR_{boot}$ \\\hline
    10\% & 15.73 & 11.14 & 10.77 & 10.31  \\                    
     5\% &  9.18 &  6.06 &  5.70 &  5.40  \\                                     
     1\% &  2.58 &  1.15 &  1.10 &  1.21  \\\hline
\end{tabular}
\end{table}

In all simulated situations, the likelihood ratio test was liberal.
Of course, this is not a proof that this is always the case. Indeed, there may
be situations where it is conservative. Simulation results presented
in the literature, however, suggest that the likelihood ratio test is often anti-conservative.
For a theoretical justification in a simple situation, let $z_{1},\ldots,z_{n}$
be a  random sample drawn from the $N(\mu,\sigma^2)$ distribution, with both $\mu$ and
$\sigma^2$ unknown. Consider the test of $\mathcal{H}_{0}\!:\mu=\mu_{0}$ versus
$\mathcal{H}_{1}\!: \mu\neq\mu_{0}$. The asymptotic likelihood ratio test rejects
$\mathcal{H}_{0}$ whenever $LR>c_\gamma$, where $c_\gamma$ is the $1-\gamma$ quantile of the 
$\chi_1^2$ distribution. Equivalently, $\mathcal{H}_{0}$ is rejected when 
$\sqrt{n}|\overline{z} - \mu_{0}|/\widehat{\sigma}>k(\gamma,n),$ where 
$\overline z = \sum_{i=1}^n z_i/n$, $\widehat{\sigma}^2 = \sum_{i=1}^{n}(z_{i}-\overline{z})^2/(n-1)$
and $k(\gamma,n) = \sqrt{(\exp(-c_{\gamma}/2)^{-2/n} - 1)(n-1)}$. Table \ref{tab8:2}
shows the true levels of the likelihood ratio test, i.e. $\Pr(LR > c_{\gamma})$ evaluated at 
$\mathcal{H}_{0}$, for different values of $n$ and $\gamma$. Notice that, even in this simple situation, 
the likelihood ratio test is liberal when the sample is not large, in agreement with 
simulation results presented elsewhere. See, for instance, Rieck and 
Nedelman~(1991, Table 4) and Cordeiro et al.~(1995).

\begin{table}[htp]
\renewcommand{\arraystretch}{1.12}
\caption{True level; normal distribution.}\label{tab8:2}
\begin{tabular}{c ccc}\hline
&\multicolumn{3}{c}{$\gamma$}\\\cline{2-4}
$n$ &   1\%  & 5\%  & 10\%   \\ \hline
5   &   2.91 & 9.79 &  16.54 \\      
8   &   1.97 & 7.64 &  13.72 \\      
12  &   1.58 & 6.64 &  12.35 \\       
20  &   1.32 & 5.93 &  11.35 \\
50  &   1.12 & 5.36 &  10.52 \\\hline
\end{tabular}                                                                                                                                                                                                                                                                  
                                                                                                                                                                                                                                                                                              \end{table}

\section{An application}\label{application}

We shall now turn to an empirical application that employs real data. We consider
the investigation made by Lepadatu et al.~(2005) on metal extrusion
die lifetime. As noted by the authors (p.\ 38), ``the estimation of tool life (fatigue
life) in the extrusion operation is important for scheduling tool changing
times, for adaptive process control and for tool cost evaluation.''
They also note (p.\ 39) that ``die fatigue cracks are caused by the repeat application
of loads which individually would be too small to cause failure.'' According
to them, current research aims at describing the whole fatigue process by
focusing on the analysis of crack propagation from very small initial defects.
It is noteworthy that fatigue failure due to propagation of an initial
crack was the main motivation for the Birnbaum--Saunders distribution.

In Section~\ref{introduction}, we explained that the interest lies in modeling the
die lifetime ($T$) in the metal extrusion process, which is mainly determined
by its material properties and by the stresses under load. The extrusion
die is exposed to high temperatures, which can also be damaging. 
The covariates are the friction
coefficient ($x_1$), the angle of the die ($x_2$) and work temperature
($x_3$). Consider a regression model which also includes interaction
effects, i.e.,
\begin{equation}\label{model1}
y_{i} = \beta_{0} + \beta_{1}x_{1i} + \beta_{2}x_{2i} + \beta_{3}x_{3i} + \beta_{4}x_{1i}x_{2i}
        + \beta_{5}x_{1i}x_{3i} + \beta_{6}x_{2i}x_{3i} + \varepsilon_{i},
\end{equation}
where $y_{i} = \log(T_{i})$ and $\varepsilon_{i}\sim\mathcal{SN}(\alpha, 0, 2)$,
$i = 1,2,\ldots,n$. There are only 15 observations ($n=15$), and we want
make inference on the significance of the interaction effects, i.e., we wish to
test $\mathcal{H}_{0}\!: \beta_{4} = \beta_{5} = \beta_{6} = 0$.
The likelihood ratio test statistic ($LR$) equals $6.387$ ($p$-value 0.094),
and the two corrected test statistics are $LR_{b} = 4.724$ ($p$-value 0.193) and
$LR_{b}^{*} = 4.492$ ($p$-value 0.213). The $p$-value of the bootstrap-based likelihood ratio
test is 0.276. It is noteworthy that one rejects the
null hypothesis at the 10\% nominal level when the inference is based on the
likelihood ratio test, but a different inference is reached when the modified
(Bartlett-corrected or bootstrap-based) tests are used. Recall from the previous section that the 
unmodified test is oversized when the sample is small (here, $n=15$), which leads us to mistrust
the inference delivered by the likelihood ratio test. 

We proceed by removing the interaction effects (as suggested by the three
modified tests) from Model~(\ref{model1}). We then estimate
\[
y_{i} = \beta_{0} + \beta_{1}x_{1i} + \beta_{2}x_{2i} + \beta_{3}x_{3i} + \varepsilon_{i},
\]
$i=1,\dots,15$. The point estimates are (standard errors in parentheses): 
$\widehat{\beta}_{0} = 5.9011\, (0.488)$, $\widehat{\beta}_{1} = 0.7917\, (1.777)$,
$\widehat{\beta}_{2} = 0.0098\, (0.012)$, $\widehat{\beta}_{3} = 0.0052\, (0.001)$
and $\widehat{\alpha} = 0.1982\, (0.036)$.
The null hypothesis  $\mathcal{H}_{0}\!: \beta_{3} = 0$ is strongly rejected
by the four tests (unmodified and modified) at the usual significance levels.
All tests also suggest the individual and joint exclusions of $x_1$ and $x_2$
from the regression model. 
We thus end up with the reduced model
\[
y_{i} = \beta_{0} + \beta_{3}x_{3i} + \varepsilon_{i}, 
\]
$i=1,\ldots,15$. The point estimates are (standard errors in parentheses): 
$\widehat{\beta}_{0} = 6.2453\, (0.326)$, $\widehat{\beta}_{3} = 0.0052\, (0.001)$
and $\widehat{\alpha} = 0.2039\, (0.037)$.

We now return to Model~(\ref{model1}) and test $\mathcal{H}_{0}\!: \beta_1 =
\beta_2 = \beta_{4} = \beta_{5} = \beta_{6} = 0$ (exclusion of all variables
but $x_3$). The null hypothesis is not rejected at the 10\% nominal level
by all tests, but we note that the corrected tests yield considerably
larger $p$-values. The test statistics are $LR = 7.229$, $LR_b = 5.610$ and
$LR_b^* = 5.417$, the corresponding $p$-values being 0.204, 0.346 and
0.367; the $p$-value obtained from the bootstrap-based likelihood ratio test 
equals 0.484.

\section{Conclusions}\label{conclusions}

We addressed the issue of performing testing inference in Birnbaum--Saunders
regressions when the sample size is small. The likelihood ratio test can be
considerably oversized (liberal), as evidenced by our numerical results.
We derived modified test statistics whose null distributions are more
accurately approximated by the limiting null distribution than that of the
likelihood ratio test statistic. We have also considered a parametric bootstrap scheme
to obtain improved critical values and accurate $p$-values for the likelihood ratio test.
Our simulation results have convincingly shown
that inference based on the modified test statistics can be much more
accurate than that based on the unmodified statistic. The modified tests behave as reliably as
a likelihood ratio test that relies on bootstrap-based critical values, with no need of
computer intensive procedures. We recommend the use
of the following statistics: $LR_{b}$ and $LR_{b}^{*}$. The latter has the
advantage of only taking on positive values, which is desirable. We have also
presented an empirical application in which the use of the finite sample
adjustment proposed in this paper can lead to inferences that are different
from the ones reached based on first order asymptotics.

\section*{Acknowledgments}

We gratefully acknowledge grants from FAPESP and CNPq. We thank an anonymous referee
for helpful comments that led to several improvements in this paper.

\appendix
{\small
\section*{Appendix}
From (\ref{lr2}), we have
\[
\epsilon_{p+1} = \sum_{r,s,t,u=1}^{p+1}\lambda_{rstu} - \sum_{r,s,t,u,v,w=1}^{p+1}\lambda_{rstuvw}.
\]               
Note that $\sum_{r,s,t,u=1}^{p+1}\lambda_{rstu}$ can be written as
$\sum_{r,s,t,u=1}^{p}\lambda_{rstu}$ plus terms in which at least
one subscript equals $\alpha$. It follows from the orthogonality between $\alpha$ and $\bsy{\beta}$
that several terms
equal zero. The non-zero terms are
$\sum_{r,s=1}^{p}\lambda_{rs\alpha\alpha}$, $\sum_{t,u=1}^{p}\lambda_{\alpha\alpha tu}$
and $\lambda_{\alpha\alpha\alpha\alpha}$. Also,
$\sum_{r,s,t,u,v,w=1}^{p+1}\lambda_{rstuvw}$ is given by
$\sum_{r,s,t,u,v,w=1}^{p}\lambda_{rstuvw}$ plus the following terms:
$\sum_{r,s,t,u=1}^{p}\lambda_{rstu\alpha\alpha}$,
$\sum_{r,s,v,w=1}^{p}\lambda_{rs\alpha\alpha vw}$,
$\sum_{t,u,v,w=1}^{p}\lambda_{\alpha\alpha tuvw}$,
$\sum_{r,s=1}^{p}\lambda_{rs\alpha\alpha\alpha\alpha}$,
$\sum_{t,u=1}^{p}\lambda_{\alpha\alpha tu\alpha\alpha}$,
$\sum_{v,w=1}^{p}\lambda_{\alpha\alpha\alpha\alpha vw}$ and
$\lambda_{\alpha\alpha\alpha\alpha\alpha\alpha}$.
Here, we present the derivations of $\sum_{r,s,t,u=1}^{p}\lambda_{rstu}$
and $\sum_{v,w=1}^{p}\lambda_{\alpha\alpha\alpha\alpha vw}$.
The other terms can be obtained in a similar fashion.

Note that $\sum_{r,s,t,u=1}^{p}\lambda_{rstu} =
(1/4)\sum_{r,s,t,u=1}^{p}\kappa^{rs}\kappa^{tu}\kappa_{rstu}.$
Inserting the cumulants given in Section~\ref{bartlett} into this expression we have
\begin{align*}
\sum_{r,s,t,u=1}^{p}\lambda_{rstu} &=
\frac{1}{4}\sum_{r,s,t,u=1}^{p}\kappa^{rs}\kappa^{tu}\Biggl\{\psi_{2}(\alpha)\sum_{i=1}^{n}x_{ir}x_{is}x_{it}x_{iu}\Biggr\}\\
&= \frac{\psi_{2}(\alpha)}{4}\sum_{i=1}^{n}\sum_{r,s,t,u=1}^{p}x_{ir}\kappa^{rs}x_{is}x_{it}\kappa^{tu}x_{iu}\\
&= \frac{\psi_{2}(\alpha)}{4}\sum_{i=1}^{n}\Biggl\{\sum_{r,s=1}^{p}x_{ir}\kappa^{rs}x_{is}\Biggr\}
\Biggl\{\sum_{t,u=1}^{p}x_{it}\kappa^{tu}x_{iu}\Biggr\}\\
&= \frac{\psi_{2}(\alpha)}{4}\sum_{i=1}^{n}\bigl(\bsy{x}_{i}^{\top}\bsy{K}^{\bsy{\beta}\bsy{\beta}}\bsy{x}_{i}\bigr)
\bigl(\bsy{x}_{i}^{\top}\bsy{K}^{\bsy{\beta}\bsy{\beta}}\bsy{x}_{i}\bigr)
= \frac{\psi_{2}(\alpha)}{4}\sum_{i=1}^{n}\bigl(\bsy{x}_{i}^{\top}\bsy{K}^{\bsy{\beta}\bsy{\beta}}\bsy{x}_{i}\bigr)^2,
\end{align*}
where $\bsy{x}_{i}^{\top} = (x_{i1}, x_{i2}, \ldots,x_{ip})$ represents the
$i$th row of $\bsy{X}$ and
$\bsy{K}^{\bsy{\beta}\bsy{\beta}}=\bsy{K}(\bsy{\beta})^{-1}
= 4(\bsy{X}^{\top}\!\bsy{X})^{-1}/\psi_{1}(\alpha)$ represents
the inverse of Fisher's information matrix for $\bsy{\beta}$.
Therefore,
\[
\sum_{r,s,t,u=1}^{p}\lambda_{rstu} =
\frac{4\psi_{2}(\alpha)}{\psi_{1}(\alpha)^2}\sum_{i=1}^{n}\bigl\{\bsy{x}_{i}^{\top}(\bsy{X}^{\top}\!\bsy{X})^{-1}
\bsy{x}_{i}\bigr\}^2.
\]
Note that $z_{ii} = \bsy{x}_{i}^{\top}(\bsy{X}^{\top}\!\bsy{X})^{-1}\bsy{x}_{i}$ is the
$i$th diagonal element of $\bsy{Z}_{\!d}$ given in Section~\ref{bartlett}. Hence,
\[
\sum_{r,s,t,u=1}^{p}\lambda_{rstu} = \frac{4\psi_{2}(\alpha)}{\psi_{1}(\alpha)^2}\sum_{i=1}^{n}z_{ii}^2
= \frac{4\psi_{2}(\alpha)}{\psi_{1}(\alpha)^2}{\rm tr}(\bsy{Z}_{\!d}^{(2)}).
\]

From
$\sum_{v,w=1}^{p}\lambda_{\alpha\alpha\alpha\alpha vw} = (1/4)
(\kappa^{\alpha\alpha})^2\kappa_{\alpha\alpha\alpha}\sum_{v,w=1}^{p}
\kappa^{vw}\kappa_{\alpha vw}$, 
we obtain
\begin{align*}
\sum_{v,w=1}^{p}\lambda_{\alpha\alpha\alpha\alpha vw} &=
\frac{\alpha^4}{4n^2}\frac{5n}{2\alpha^3}\sum_{v,w=1}^{p}\kappa^{vw}\Biggl\{
\frac{2+\alpha^2}{\alpha^3}\sum_{i=1}^{n}x_{iv}x_{iw}\Biggr\}\\
&= \frac{5(2+\alpha^2)}{8n\alpha^2}\sum_{i=1}^{n}\Biggl\{\sum_{v,w=1}^{p}x_{iv}\kappa^{vw}x_{iw}\Biggr\}
= -\frac{5(2+\alpha^2)}{8n\alpha^2}\sum_{i=1}^{n}\bigl(\bsy{x}_{i}^{\top}\bsy{K}^{\bsy{\beta}\bsy{\beta}}\bsy{x}_{i}\bigr)\\
&=-\frac{5(2+\alpha^2)}{2n\alpha^2\psi_{1}(\alpha)}\sum_{i=1}^{n}\bigl\{\bsy{x}_{i}^{\top}(\bsy{X}^{\top}\!\bsy{X})^{-1}\bsy{x}_{i}\bigr\}\\
&= -\frac{5(2+\alpha^2)}{2n\alpha^2\psi_{1}(\alpha)}\sum_{i=1}^{n}z_{ii} = -\frac{5(2+\alpha^2)}{2n\alpha^2\psi_{1}(\alpha)}{\rm tr}
(\bsy{Z}_{\!d}) = -\frac{5(2+\alpha^2)p}{2n\alpha^2\psi_{1}(\alpha)}.
\end{align*}


{\small
\begin{thebibliography}{99}

\bibitem{AbellBrase94}
Abell, M.~L. and Braselton, J.~P. (1994).
\newblock {\em The Maple V Handbook\/}.
\newblock AP Professional, New York.                                          

\bibitem{Bartlett37}
Bartlett, M.~S. (1937).
\newblock Properties of sufficiency and statistical tests.
\newblock {\em Proceedings of the Royal Society A\/}, {\bf 160}, 268--282.

\bibitem{BSa}
Birnbaum, Z.~W. and Saunders, S.~C. (1969).
\newblock A new family of life distributions.
\newblock {\em Journal of Applied Probability\/}, {\bf 6}, 319--327.

\bibitem{Cordeiro95}
Cordeiro, G. M., Cribari--Neto, F., Aubin, E. C. Q. and Ferrari, S. L. P. (1995).
\newblock Bartlett corrections for one-parameter exponential family models.
\newblock {\em Journal of Statistical Computation and Simulation},
{\bf 53}, 211--231. 

\bibitem{CoxReid87}
Cox, D.~R. and Reid, N. (1987).
\newblock Parameter orthogonality and approximate conditional inference.
\newblock {\em Journal of the Royal Statistical Society B\/}, {\bf
  40}, 1--39.

\bibitem{CribariNetoCordeiro96}
Cribari--Neto, F. and Cordeiro, G.~M. (1996).
\newblock On Bartlett and Bartlett-type corrections.
\newblock {\em Econometric Reviews\/}, {\bf 15}, 339--367.
 
\bibitem{CNZarcos03}
Cribari--Neto, F. and Zarkos, S. (2003).
\newblock Econometric and statistical computing using Ox.
\newblock {\em Computational Economics\/}, {\bf 21}, 277--295.

\bibitem{DesmondeA}
Desmonde, A.~F. (1985).
\newblock Stochastic models of failure in random environments.
\newblock {\em Canadian Journal of Statistics\/}, {\bf 13}, 171--183.

\bibitem{DcK}
Doornik, J.~A. (2006).
\newblock {\em An Object-Oriented Matrix Language -- Ox 4\/}.
\newblock Timberlake Consultants Press, London.
\newblock 5th ed.

\bibitem{ET}
Efron, B. and Tibshirani, R.~J. (1993).
\newblock {\em An Introduction to the Bootstrap\/}.
\newblock Chapman and Hall, New York.

\bibitem{GLP}
Galea, M., Leiva, V. and Paula, G.~A. (2004).
\newblock Influence diagnostics in log-Birnbaum--Saunders regression models.
\newblock {\em Journal of Applied Statistics\/}, {\bf 31}, 1049--1064.

\bibitem{GR65}
Gradshteyn, I.~S. and Ryzhik, I.~M. (2007).
\newblock {\em Table of Integrals, Series, and Products\/}.
\newblock Academic Press, New York.

\bibitem{JohnsonetalV2}
Johnson, N., Kotz, S. and Balakrishnan, N. (1995).
\newblock {\em Continuous Univariate Distributions -- Volume 2\/}, 2nd ed.
\newblock Wiley, New York.

\bibitem{KNBala}
Kundu, D., Kannan, N. and Balakrishnan, N. (2008).
\newblock On the function of Birnbaum--Saunders distribution and associated inference.
\newblock {\em Computational Statistics and Data Analysis\/}, {\bf 52}, 2692--2702.
                  
\bibitem{Lawley}
Lawley, D. (1956).
\newblock A general method for approximating to the distribution of likelihood
  ratio criteria.
\newblock {\em Biometrika\/}, {\bf 43}, 295--303.

\bibitem{LBPG}
Leiva, V., Barros, M.~K., Paula, G.~A. and Galea, M. (2007).
\newblock Influence diagnostics in log-Birnbaum--Saunders regression models
  with censored data.
\newblock {\em Computational Statistics and Data Analysis\/}, {\bf 51},
  5694--5707.

\bibitem{LCNV}
Lemonte, A.~J., Cribari--Neto, F. and Vasconcellos, K.~L.~P. (2007).
\newblock Improved statistical inference for the two-parameter
  Birnbaum--Saunders distribution.
\newblock {\em Computational Statistics and Data Analysis\/}, {\bf 51},
  4656--4681.

\bibitem{LSCN}
Lemonte, A.~J., Simas, A.~B. and Cribari--Neto, F. (2008).
\newblock Bootstrap-based improved estimators for the two-parameter
  Birnbaum--Saunders distribution.
\newblock {\em Journal of Statistical Computation and Simulation\/}, {\bf 78},
  37--49.

\bibitem{Lepadato-et-al}
Lepadatu, D., Kobi, A., Hambli, R. and Barreau, A. (2005).
\newblock Lifetime multiple response optimization of metal extrusion die.
\newblock {\em Proceedings of the Annual Reliability and Maintainability Symposium\/},
37--42.

\bibitem{McCarter}
McCarter, K.~S. (1999).
\newblock {\em Estimation and Prediction for the Birnbaum--Saunders
  Distribution Using Type-II Censored Samples, With a Comparison to the Inverse
  Gaussian Distribution\/}.
\newblock Ph.D.~dissertation, Kansas State University.

\bibitem{Nelson}
Nelson, W. (1990).
\newblock {\em Accelerated Testing, Statistical Models, Test Plans and Data
  Analysis\/}.
\newblock Wiley, New York.

\bibitem{Pressetal}
Press, W.~H., Teulosky, S.~A., Vetterling, W.~T. and Flannery, B.~P. (1992).
\newblock {\em Numerical Recipes in C: The Art of Scientific Computing\/}, 2nd ed.
\newblock Prentice Hall, London.

\bibitem{CoreTeam}
{R Development Core Team} (2006).
\newblock {\em R: A Language and Environment for Statistical Computing\/}.
\newblock R Foundation for Statistical Computing, Vienna, Austria.

\bibitem{Rieck89}
Rieck, J.~R. (1989).
\newblock {\em Statistical Analysis for the Birnbaum--Saunders Fatigue Life
  Distribution\/}.
\newblock Ph.D.\ dissertation, Clemson University.

\bibitem{RiekNedelman}
Rieck, J.~R. and Nedelman, J.~R. (1991).
\newblock A log-linear model for the Birnbaum--Saunders distribution.
\newblock {\em Technometrics\/}, {\bf 33}, 51--60.

\bibitem{Saunders}
Saunders, S.~C. (1974).
\newblock A family of random variables closed under reciprocation.
\newblock {\em Journal of the American Statistical Association\/}, {\bf 69},
  533--539.

\bibitem{Tisionas}
Tisionas, E.~G. (2001).
\newblock Bayesian inference in Birnbaum--Saunders regression.
\newblock {\em Communications in Statistics -- Theory and Methods\/}, {\bf 30},
  179--193.

\bibitem{XiWei}
Xie, F.~C. and Wei, B.~C. (2007).
\newblock Diagnostics analysis for log-Birnbaum--Saunders regression models.
\newblock {\em Computational Statistics and Data Analysis\/}, {\bf 51},
  4692--4706.

\end{thebibliography}
}
\end{document}